\documentclass[12pt,preprint]{aastex}

\usepackage{graphicx}
\usepackage{natbib}
\def\old#1{}

\def\oldmath#1{}

\def\abin{a_{\rm bin}}
\def\rclose{R_{\rm min}}
\def\acore{a_{\rm c}}

\def\ma{m_1}
\def\mb{m_2}
\def\va{v_1}
\def\vb{v_2}
\def\vej{v_{\rm ej}}
\def\tlife{t_{\rm ms}}
\def\Mbh{M_\bullet}
\def\Msolar{M$_\odot$}
\def\msun{M$_\odot$}
\def\mathMsolar{{\rm M}_\odot}
\def\facR{f_R}
\def\Sgr{Sgr~A$^\ast$}
\def\GC{Galactic Center}
\def\kms{km~s$^{-1}$}

\begin{document}

\title{Hypervelocity Stars: Predicting the Spectrum of Ejection Velocities}

\author{Benjamin C. Bromley}
\affil{Department of Physics, University of Utah, 
\\ 115 S 1400 E, Rm 201, Salt Lake City, UT 84112}
\email{bromley@physics.utah.edu}

\author{Scott J. Kenyon}
\affil{Smithsonian Astrophysical Observatory,
\\ 60 Garden St., Cambridge, MA 02138}
\email{skenyon@cfa.harvard.edu}

\author{Margaret J. Geller}
\affil{Smithsonian Astrophysical Observatory,
\\ 60 Garden St., Cambridge, MA 02138}
\email{mgeller@cfa.harvard.edu}

\author{Elliott Barcikowski}
\affil{Department of Physics, University of Utah,
\\ 115 S 1400 E, Rm 201, Salt Lake City, UT 84112}
\email{elliottb@physics.utah.edu}

\author{Warren R. Brown}
\affil{Smithsonian Astrophysical Observatory,
\\ 60 Garden St., Cambridge, MA 02138}
\email{wbrown@cfa.harvard.edu}

\author{Michael J. Kurtz}
\affil{Smithsonian Astrophysical Observatory,
\\ 60 Garden St., Cambridge, MA 02138}
\email{mkurtz@cfa.harvard.edu}

\begin{abstract}
The disruption of binary stars by the tidal field of the black hole in
the Galactic Center can produce the hypervelocity stars observed in
the halo.  We use numerical models to simulate the full spectrum of
observable velocities of stars ejected into the halo by this binary
disruption process. Our model includes a range of parameters for
binaries with 3-4 M$_\odot$ primaries, consideration of radial orbits
of the ejected stars through an approximate mass distribution for the
Galaxy, and the impact of stellar lifetimes. We calculate the spectrum
of ejection velocities and reproduce previous results for the mean
ejection velocity at the Galactic center. The model predicts that the
full population of ejected stars includes both the hypervelocity stars
with velocities large enough to escape from the Galaxy and a
comparable number of ejected, but bound, stars of the same stellar
type. The predicted median speeds of the population of ejected stars
as a function of distance in the halo are consistent with current
observations. Combining the model with the data also shows that
interesting constraints on the properties of binaries in the Galactic
Center and on the mass distribution in the Galaxy can be obtained even
with modest samples of ejected stars.

\end{abstract}


\maketitle

\section{Introduction}

Hypervelocity stars (HVSs) are objects moving with speeds sufficient 
to escape the gravitational influence of the Galaxy. Hills
\citeyearpar{hil88} predicted their existence, and Brown et al.\
\citeyearpar{broetal05} discovered the first HVS in a survey of 
blue stars within the Galactic halo. We now know of seven HVSs
\citep{ede05,hir05,broetal06a,broetal06b}.  One HVS is a subdwarf 
O star at a distance of $\sim$ 20 kpc from the \GC\ \citep{hir05}. 
The other HVSs are probably B-type main sequence stars with galactocentric 
distances of 50--100 kpc. The five late B stars in this group have masses 
$m \sim$ 3-4 \msun\ \citep{broetal06a,broetal06b}. The early B-type
HVS is more massive, $m \sim$ 8 \msun\ \citep{ede05}.

Only two proposed models plausibly explain the origin of the observed
population of Galactic HVSs. In the single black hole model, a binary system 
strays too close to \Sgr, the 3.5 million solar-mass black hole in the \GC\ 
\citep{hil88,hil92,yutre03,gouqui03}.  This encounter tears the binary apart, 
leading to the capture of one star and the high-speed ejection of its partner 
\citep{hil88}.  For typical ejection velocities of $\sim$ 2000 km s$^{-1}$,
the HVS can reach galactocentric distances of $\sim$ 50 kpc in $\sim$ 25 Myr
\citep{hil88,hil92,yutre03,gouqui03,broetal06b,perhopale06}, roughly the 
lifetime of a 9 \msun\ star \citep{schaetal92, schaetal93}.

Other HVS models involve three-body interactions between a single star
and a binary black hole. Random encounters between stars near the \GC\
and a binary black hole can lead to an HVS with a typical ejection
velocity of $\sim$ 2000 km s$^{-1}$
\citep{qui96,yu02,yutre03,gua05,ses06}. Alternatively, the in-spiral of an
intermediate mass black hole with $m \sim 10^3-10^4$ \msun\ can eject
stars from the \GC\ \citep[e.g.,][]{han03}.  Some of these encounters
will produce HVSs with ejection velocities of $\sim$ 2000 km s$^{-1}$
\citep{lev05,BauGuaPor06,ses06}. Although binary black holes with
circular orbits and in-spiraling black holes produce roughly isotropic
distributions of HVSs on the sky, eccentric binary black holes can
produce anisotropic space distributions of HVSs
\citep{hol05,BauGuaPor06,ses06}.

Here, we use numerical simulations to predict the distribution of
observable radial velocities of HVSs ejected by a single black hole at
the \GC. Our results identify a new population of ejected stars of the
same stellar type as the HVSs but more slowly moving.  Ongoing halo
surveys \citep{broetal06b} show evidence of this population.  The
distribution of observable radial velocities among these stars may
yield information about the nature of the progenitor binaries and can
serve as a probe of the distribution of mass throughout the
Galaxy. Even the small number of known HVSs can constrain the
distribution of mass in the Milky Way \citep[see also][]{gneetal05}.

We describe the basic model and the numerical simulations in \S2,
compare the numerical results with observations in \S3, and conclude
with a discussion and summary in \S4.

\section{The Model}

\subsection{Overview}

The goal of this study is to develop methods for predicting the
observable velocity distribution of HVSs in the Galactic halo. We
derive velocity distributions appropriate for the single black hole at
the \GC; we do not consider models for ejections from binary black
holes \citep[e.g.,][]{yutre03,gua05,lev05}.  Previous analytical and
numerical analyses have concentrated on the production of HVSs from
the interactions of binary stars with the massive black hole in the
\GC\ \citep{hil88,hil92,yutre03,gouqui03,perhopale06}.  However, the
observable distribution of HVSs also depends on the mass distribution
in the \GC\ and on the gravitational potential of the Milky Way
\citep[e.g.,][]{gneetal05}.  Here, we adopt a simple prescription for
the Galactic mass distribution, integrate the orbits of stars ejected
from the \GC, and derive a first approximation to the full observable
phase-space distribution of ejected stars on radial orbits in the
halo. Comparisons between observations of halo stars with the model
predictions then yield constraints on the mass function of binaries in
the \GC\ and the Galactic potential.

We simulate the ejection of stars from the \GC\ in three steps. Following 
\citet{hil88}, we first integrate orbits of binary stars passing by \Sgr\ 
to quantify ejection probabilities and velocities. Then we use a Monte Carlo 
code based on semianalytical approximations for rapid generation of simulated 
catalogs of ejected stars in the \GC. Finally we integrate the orbits of these 
stars in the Galactic potential to derive how they populate the Galaxy's halo.
To generate observable samples, the simulation incorporates the stellar main 
sequence lifetime from published stellar evolution calculations.  In this first 
study, we focus on binaries with the full range of possible initial separations 
and with primary star masses of 3--4~\Msolar, corresponding to halo B stars 
targeted in ongoing radial velocity surveys \citep{broetal06a,broetal06b}.

\subsection{Simulations of Binary Interactions with the Central 
Black Hole}

To follow the evolution of a binary as it interacts with a massive
black hole, we use a sixth-order, symplectic integrator \citep{yos90}.
\citet{broken06} describe tests of this code in the context of a
simulation of planet formation \citep[see also][]{kenbro06}.  For the
three-body simulations here, the algorithm restricts energy errors to
less than one part in $10^9$.  Our numerical trials start with
randomly oriented binary stars, launched toward the black hole from a
distance of several thousand AU, much larger than the binary
separation (generally less than a few AU).  As in \citet{hil88}, we
assume that the initial approach speed of the binary center of mass is
250~km s$^{-1}$, and that the binary star orbit has negligible
eccentricity.  Variations in the initial approach speed have little
impact on the results. Although the binary eccentricity can affect the
outcome of an ejection event, stars in eccentric binaries spend most
of their time at separations larger than the orbital semimajor
axis. Thus, results for eccentric binaries generally mimic results for
circular binaries with appropriately larger semimajor axes.  Table
\ref{tab:modelparms} lists the ranges of other input parameters used
in these calculations.

Our simulations reproduce published numerical results and they are 
consistent with analytical expectations for the production of HVSs 
\citep{hil88,hil92,yutre03,gouqui03}.  The high-speed ejection of a 
binary member from an encounter with a massive black hole depends on 
the binary's semimajor axis $\abin$, the closest-approach distance 
between the binary and the black hole $\rclose$, and the masses of 
the three bodies --- $\Mbh$ is the black hole mass, and $\ma$ and 
$\mb$ are the masses of the primary and secondary binary partners, 
respectively.  For encounters which produce an ejected star, our
numerical simulations yield \citep[see also][]{hil88}
\begin{equation}\label{eq:vej}
\vej = 1760 
\left(\frac{\abin}{\rm 0.1\ AU}\right)^{-1/2} 
       \left(\frac{\ma + \mb}{2 \mathMsolar}\right)^{1/3}
\left(\frac{\Mbh}{3.5 \times 10^6 \mathMsolar}\right)^{1/6}
       \facR \ \   {\rm km~s^{-1}} 
\end{equation}
where the factor $\facR$ is of order unity, and serves to tune the 
ejection speed in accord with numerical simulations 
\citep[as in Fig.~3 of][]{hil88}. To approximate Hills' results, 
we derive a quintic polynomial fit for $\facR$
\begin{eqnarray}\label{eq:facR}
\nonumber
    \facR & = & 0.774+(0.0204+(-6.23\times 10^{-4}
       +(7.62\times 10^{-6}+ \\
& &  
\nonumber
(-4.24\times 10^{-8}
        +8.62\times 10^{-11}D)D)D)D)D,
\end{eqnarray}
where Hills' parameter $D$ is defined in equation~(\ref{eq:D}) below.
For an unequal mass binary, the ejection speeds
of the primary and secondary are 
\begin{equation}\label{eq:vavb}
\va = \vej \left(\frac{2\mb}{\ma+\mb}\right)^{1/2}  \ \ \ {\rm and} \ \ \ 
\vb = \vej \left(\frac{2\ma}{\ma+\mb}\right)^{1/2} ,
\end{equation}
respectively. 

We confirm that both stars are rarely ejected during an encounter
\citep{hil88}.  Furthermore, for the types of binary orbits and mass
ranges we consider, the primary and secondary have an approximately
equal chance of ejection.  The key factors in the ejection are the
orbital phase and orientation of the orbital plane when the binary
comes under the grip of the black hole's tidal field.  Our simulations
suggest that if a binary is disrupted, then the star on the orbit
closest to the black hole tends to be captured; its partner is ejected.
When there is a significant difference between primary and secondary masses, 
the primary is more likely to be ejected. We can understand this preference 
in the limit of a very large mass ratio, where the primary follows its 
unbound Keplerian orbit about the black hole and the secondary may get 
captured.  However, for the mass ratios and orbital configurations which 
produce the ejection speeds of interest here, our simulations suggest that 
we may neglect any preference in the ejection of the primary star.

The ejection speeds in equations~(\ref{eq:vej}) and (\ref{eq:vavb})
represent averages, and are the theoretical speeds at infinite
distance from the black hole in the absence of other gravitational
sources.  Figure~\ref{fig:vdisp} illustrates our calculated
distribution of ejection speeds for equal-mass binaries after
encountering \Sgr, with $\Mbh$ = $3.5\times 10^6$~\Msolar.  The
distribution has a pronounced peak at $\sim$ 2800 km s$^{-1}$ and
broad wings extending to $\sim$ 1000 km s$^{-1}$ and $\sim$ 4000 km
s$^{-1}$.  The solid line in Figure~\ref{fig:vdisp} shows that a
Gaussian with $v_{avg} \sim$ 2600 km s$^{-1}$ and $\sigma_v \sim$ 0.2
$v_{avg} \sim$ 500 km~s$^{-1}$ provides a reasonable characterization
of the numerical results.  Repeated trials for a wide range of binary
configurations indicate that the Gaussian model generally gives an
adequate description of the ejection speed distribution.  We use this
simple Gaussian model as input to the Monte Carlo simulations of the
predicted distribution of speeds as a function of distance from the
Galactic Center.  Our simulations confirm that the final distribution
of ejected stars in the halo is insensitive to our particular choice
of $\sigma_v$ within reasonable limits.

The shape of the velocity distribution in Figure~\ref{fig:vdisp}
reflects the detailed dependence of an ejection event on the orbital
phase and orientation of the binary orbit relative to the black hole.
We speculate that the extended tails of the distribution correspond to
infrequent encounters where the binary angular momentum is either
strongly aligned or counteraligned with its center-of-mass angular
momentum about the black hole. Otherwise, encounters yield velocities
near the mean, causing a peak in the distribution near $v_{avg}$.  A
study of a large ensemble of simulations may confirm this speculation
and is beyond the scope of this initial study.

Given values of $\abin$ and $\rclose$, we derive the probability that
a particular interaction leads to the ejection of a star. Following
\citet{hil88}, we calculate a dimensionless quantity,
\begin{equation}\label{eq:D}
D = 
\left(\frac{\rclose}{\abin}\right) 
       \left[\frac{2 \Mbh}{10^6 (\ma + \mb)}\right]^{-1/3},
\end{equation}
and derive the probability of an ejection as
\begin{equation}
\label{eq:PE}
P_{ej} \approx 1-D/175
\end{equation} 
for $0 \le D \le 175$. For $D > 175$, $\rclose \gg \abin$, and the
binary does not get close enough to the black hole for an ejection.
Thus $P_{ej}$ is zero 
in this case.

Our treatment of the ejection probability does not account for the
possibility that a black hole-binary encounter will lead to a
collision between the binary partners \citep{yutre03,ginloe06}.  Our
simulations suggest that stellar collisions may be important for small
values of $\abin$, but only at the O(10\%) level. Furthermore, there
is enough uncertainty in the possible outcome of such high-speed
collisions, including rapidly rotating ejected stars or mergers
\citep{ginloe06}, that we do not consider their effect in this
work. Instead we note that equation~(\ref{eq:PE}) may modestly
overestimate the ejection probability in some cases.

Next, we calculate the broader distribution of ejection velocities that arise 
from encounters with a range of astrophysically relevant 
$\abin$ and $\rclose$ values.  
Surveys of large samples of binary systems with solar-type primary stars 
\citep[e.g.][] {abt83,hea98} suggest that the probability density function 
for binary semimajor axis is roughly
\begin{equation}\label{eq:pabin}
p(\abin) d\abin \sim d\abin/\abin
\end{equation} 
for $\abin \sim 10^{-2}-10^3$ AU. \citet{duq91} favor a Gaussian
distribution in log $P$, where $P$ is the orbital period. Here we
adopt equation~(\ref{eq:pabin}), which yields an equal number of
objects in equally spaced logarithmic bins of semimajor axis and is
consistent with the distribution of $\abin$ derived from observations
of smaller samples of binaries with O-type and B-type primary stars
\citep{gar80,kob06}.  The probability density function of closest
approach distances to \Sgr\ varies linearly with $\rclose$ as a result
of gravitational focusing \citep{hil88}. With these probabilities, we
use a Monte Carlo code to draw samples of $\abin$ and $\rclose$. We
reject some fraction of samples according to equation~(\ref{eq:PE}),
and feed the rest into equations~(\ref{eq:vej}) and (\ref{eq:vavb}) to
derive average ejection speeds. A normally distributed random number
generator then provides ejection speeds with a 20\% dispersion about
the mean (eq.~[\ref{fig:vdisp}]).  Thus, we produce catalogs of
ejection events for specific binary masses, and construct
distributions of ejection speeds.

Unlike previous studies of ejected stars from the \GC, which were
focussed on HVSs, we admit wide ranges of $\abin$ and $\rclose$ to
consider ejection events that would not necessarily yield stars
capable of escaping the Galaxy. For the binary semimajor axis, we
choose $0.05~\rm{AU} \le \abin \le 4$~AU; the lower limit is suggested
by the physical radius of the 4~\Msolar\ primary (about half this
length); the upper limit arises because we are interested in stars
with sufficient ejection speeds to populate the Galactic halo beyond
10 kpc (cf.\ eq.~[\ref{eq:vej}]). Once we set the maximum $\abin$
value, equation~(\ref{eq:PE}) gives the maximum value of $\rclose$
that could result in a non-zero ejection probability.

The semianalytical approximations in
equations~(\ref{eq:vej})--(\ref{eq:pabin}) encode the underlying
physics of ejections from the \GC. Hills' parameter $D$ and the
expression for the ejection probability in equation~(\ref{eq:PE}) are
statements that the ability of \Sgr\ to disrupt a binary depends most
strongly on the ratio between the orbital separation and the closest
approach distance; a tight binary is hard to disrupt when it does not
get close to the black hole.  However, once a binary is disrupted, the
speed of the ejected star comes mainly from the kinetic energy
available to the star from its binary orbit, which scales as $1/\abin$
(eq.~[\ref{eq:vej}]).  Finally, a broad distribution of ejection
speeds arises from the interplay between binary separation and
ejection probability. For example, a small binary separation---which
can yield a high-speed ejection---is more likely than a large $\abin$,
yet it has a low probability of producing an ejection for typical
values of $\rclose$ (eq.~[\ref{eq:pabin}]). Conversely, a large binary
separation---which leads to a low-velocity ejection---is not common;
however, should a system with large $\abin$ encounter \Sgr, then an
ejection event is highly likely.  Within realistic ranges of $\abin$ and
$\rclose$, ejection speeds can vary widely.

\subsection{The Galactic Potential and the Observed Velocity Spectrum}

To construct simulated catalogs of ejected stars that are relevant 
to HVSs searches in the Galactic halo, we next consider the effect of 
the Galactic potential. A simple parameterization of the distribution 
of mass in the Galaxy is
\begin{equation}\label{eq:massmodel}
\rho(r) = \frac{\rho_0}{\left[1+(r/\acore)^\alpha\right]},
\end{equation}
where $r$ is the distance to the \GC, $\rho_0$ is the central density,
$\acore$ is a ``core radius,'' and the index $\alpha$ is around two
\citep{bintre87}.  Unless otherwise specified, we set $\rho_0 =
1.27\times 10^4$~\Msolar~pc$^{-3}$, $\acore = 8$~pc, and $\alpha = 2$;
these choices give a mass within 10~pc of the Galactic Center of
$\sim 3\times 10^7$~\Msolar, as inferred from stellar kinematics in
the region of \Sgr\ \citep{eckgen97,schoetal03,gheetal05}, and a
circular rotation speed of 220~km~s$^{-1}$ in the Solar neighborhood
\citep{hog05}.

To produce simulated surveys, we view the \GC\ as a fountain of
ejected stars, spewing out at a steady rate.  We focus on primary
stars of 3~\Msolar\ to 4~\Msolar, because these objects are the targets 
of ongoing B~star searches. Although B~stars could be secondaries, we
assume for this work that such binaries are relatively rare. Whether
an ejected star becomes part of a simulated survey depends on 
(i) the spatial extent of the survey, (ii) the main sequence lifetime 
of the star, and (iii) the time when the star was ejected relative to 
the present. 

For the survey volume, we assign a 10~kpc inner radius and an outer
radius of 90~kpc for 3~\Msolar\ stars and 120~kpc for 4~\Msolar\
stars, consistent with the selection in the survey of
\citet{broetal06a, broetal06b}.  To derive the time since a star was
ejected, we randomly generate an age, $T$, from a uniform distribution
between zero and the main-sequence lifetime $\tlife$, which is 350~Myr
for 3~\Msolar\ stars and 160~Myr for 4~\Msolar\ stars
\citep{schaetal92,schaetal93,broetal06a}.  In doing so, we assume that
the stars' pre-main sequence phase is comparatively short, that the
travel time between the parent binary and the black hole is also
short, and that the scattering of the host binary toward \Sgr\ is a
random process which is equally likely at any time (see
\citealp{yutre03} and \citealp{perhopale06} for descriptions of
scattering mechanisms).

We use a second orbit integration code to calculate the radial
trajectory of an ejected star as it travels through the potential
generated by the mass density in equation~(\ref{eq:massmodel}). This
code takes its input from the Monte Carlo ejection-speed generator,
modified to account for the finite distance from the black hole.  We
start each ejected star's orbit near \Sgr, at a distance where the
Galactic mass model generates 5\% of the black hole mass, typically
within a parsec.  The code integrates the orbit outward from this
starting point, with a simple leapfrog (second-order symplectic)
scheme to track radius and radial velocity as a function of time. It
concurrently integrates the density (eq.~[\ref{eq:massmodel}]) using
Simpson's rule, to derive values of the mass enclosed at each radius,
a quantity needed for force evaluation.  The integrator breaks off
when the time reaches the star's randomly chosen age, $T$, or if the
star falls back out of the survey volume toward the black hole.  In
this way, we build up simulated catalogs of ejected stars within the
Galactic halo.

Figures~\ref{fig:fountain}--\ref{fig:fountainb} summarize the main results of 
these simulations. 
Figure~\ref{fig:fountain} shows the observable velocity distribution as a
function of secondary mass. The individual panels correspond to a
4~\Msolar\ primary, and secondary masses of 0.5, 1, 2 and 4~\Msolar,
respectively. In all cases, the velocity distribution of stars in the
range 10~kpc~$<~r~<~$120~kpc is peaked near 300~km~s$^{-1}$, with virtually 
no dependence on the mass of the secondary.  Instead, the location of
this peak is largely determined by the radial extent of the survey
volume and the main-sequence lifetime. In the fountain picture, many
of the slowly moving ejected stars can reach the inner halo ($r~<~40$~kpc) 
within a time $t<\tlife$, and some will even fall back toward the \GC. 
However, the ejection speeds must be faster just to reach larger radii 
within the stellar main sequence lifetime. Thus, more distant regions 
of the halo tend to hold faster-moving ejected stars. 
Figures~\ref{fig:fountain}--\ref{fig:fountainb} show this effect by 
comparing the velocity distribution peak for stars inside $r = 40$~kpc 
(blue histograms) with the distribution for stars located outside of that 
radius (magenta histograms).  The velocity distributions for nearby stars 
peak at $\sim$ 100~km~s$^{-1}$; the peak moves to 400--500~km~s$^{-1}$ for 
more distant objects.

The overall velocity distribution of primary stars is insensitive to
the mass of the secondary. In general, the stellar lifetime combined
with the Galactic potential `filter' the distribution to admit
slower-speed ejected stars which are prevalent independent of the
secondary mass. These slower-speed ejecta live long enough to populate
the inner Galactic halo in larger numbers compared to the high
velocity tail. At larger distances, only the high velocity tail is
observable as a result of the short stellar lifetimes.  The dependence
of ejection velocity on secondary mass (equations
\ref{eq:vej}--\ref{eq:vavb}) is important only for the high velocity
tail of the observed velocity distribution.  For example, the
observation of a blue star with a speed in excess of 2000~km~s$^{-1}$
requires a secondary of at least 1~\Msolar, and statistically favors
an even larger mass because the likelihood of a hypervelocity ejection
increases with the secondary mass.

Figure~\ref{fig:fountaina} illustrates the effect of the Galactic mass
density profile on speeds of ejected 4 \msun\ stars in the halo. Although
the peak of the velocity distribution is relatively independent of the 
mass model, the median velocity is sensitive to model parameters. Changing 
the index $\alpha$ in equation~(\ref{eq:massmodel}) from 1.8 to 2.5 produces
a significant variation in the median velocity of objects in the full survey 
volume (black histograms in each panel), from 270~km~s$^{-1}$ for $\alpha$ 
= 1.8 to more than 600~km~s$^{-1}$ for $\alpha$ = 2.5.  The blue and magenta 
histograms show the sensitivity of the median to distance from the \GC.
For halo stars beyond 40 kpc, the median velocity varies from 485~km~s$^{-1}$ 
for $\alpha$ = 1.8 to 750~km~s$^{-1}$ for $\alpha$ = 2.5. 

Figure~\ref{fig:fountainb} shows similar results for binaries with a primary 
mass of 3~\Msolar\ and secondary mass of 1.5~\Msolar.  Compared to the 
4~\Msolar\ cases, the histograms shift to slower ejection speeds due to
the lower luminosity and longer (350~Myr) lifetime of the lower mass stars.
At the magnitude limit of optical surveys, stars with smaller optical 
luminosities have smaller distances; thus these surveys are shallower for 
3~\Msolar\ stars than for 4~\Msolar\ stars.
 
Table \ref{tab:medvej} summarizes the results for the median radial velocities 
in Figures \ref{fig:fountaina} and \ref{fig:fountainb}. For each adopted
$\alpha$, the Table lists the core radius ($a_c$) and central mass density
($\rho_0$) needed for a mass of $\sim 3 \times 10^7$ \msun\ within 10 pc of
the \GC\ and the median velocities for two samples of ejected stars for 
3 \msun\ and 4 \msun\ primary stars. For 3 \msun\ primaries, the full survey 
volume extends from 10 kpc to 90 kpc; the halo volume extends from 30 kpc
to 90 kpc. For 4 \msun\ primaries, the full survey volume extends from 10 kpc 
to 120 kpc; the halo volume extends from 40 kpc to 120 kpc.

In addition to producing smaller median observable speeds for lower
mass stars, the main sequence lifetimes set the maximum stellar mass
for HVSs.  For primary stars ejected with velocity $v_{ej}$ from the
\GC, Table \ref{tab:dmax} lists the maximum distance $d_{max}$ the
star can reach during its main sequence lifetime. Because the travel
time from the \GC\ to the Galactic halo, $t \propto m^{-1/3}$, has a
much weaker mass dependence than the main sequence lifetime, $\tlife
\propto m^{-2}$, massive stars ejected from the \GC\ cannot reach
large galactocentric distances.  Furthermore, high ejection speeds
require a small binary semimajor axis (eq.~[\ref{eq:vej}]); larger,
more massive stars have a larger minimum $a_{bin}$ than smaller, less
massive stars, based on an assessment of Roche lobe overflow
\citep{egg83}. Thus, less massive stars have comparable maximum
ejection velocities -- and hence larger $d_{max}$, given enhanced
stellar lifetimes -- than more massive stars.  Based on the relative
ejection velocities, stellar lifetimes, and the time needed for any
star to reach the central black hole, we suggest that the maximum mass
for an HVS in the outer halo is $\sim$ 12 \msun\
(Table~\ref{tab:dmax}).  With large surveys to greater depth, this
mass limit may provide an observational test of these models.

Figures~\ref{fig:fountain}--\ref{fig:fountainb} show the breadth of
speeds that ejected stars can have once they have traveled into the
Galactic halo, from negative values --- corresponding to infall --- to
hypervelocities.  Clearly, HVSs have slower-moving counterparts of the
same stellar type that originated near the \GC\ and carry information
about the physical process which led to their ejection.  To illustrate
this feature of the simulations in more detail, Figure~\ref{fig:vesc}
shows the fraction of escaping stars (with velocities exceeding the
local escape velocity\footnote{ We calculate the escape velocity from
the mass distribution in equation~(\ref{eq:massmodel}) for a Galaxy
with an outer radius of 120 kpc. The results are nearly identical for
outer radii of 250 kpc.}) as a function of distance from the \GC. For
3 \msun\ primaries, the fraction increases roughly linearly with
radius and is almost 50\% at $\sim$ 90 kpc, the limit of current
surveys for these stars. The fraction of escaping stars is much larger
for 4 \msun\ primaries and is 100\% at 80--120 kpc. In an ensemble of
3--4 \msun\ ejected stars, the fraction of escaping stars is $\sim$
50\%, suggesting that observations should reveal roughly comparable
numbers of unbound HVSs and ejected, but bound, halo stars with lower
observed radial velocities.  To test this and other aspects of the
calculation, we now compare our results with observed velocity
distributions of stars in the Galactic halo.

\section{Comparing the model with observations}

The model for HVSs has several observable and testable
consequences. To produce an observable ensemble of ejected stars, the
\GC\ must have a significant population of young binary stars close
enough to the central black hole.  

Current observations provide broad support for many young stars at the
\GC\ \citep[e.g.,][]{mar05}. 
Near \Sgr, high spatial resolution observations reveal $\sim$ 100 or more 
young OB stars and many evolved Wolf-Rayet stars
\citep{ghe03,schoetal03,eis05,tan05,pau06}. 
Within 25--50~pc of \Sgr, there are several distinct star-forming regions,
including the Arches cluster, containing thousands of stars with
masses $\gtrsim$ 3--4 \msun\ \citep{fig99,naj04,sto05}. 
If the binary fraction of the OB population near the \GC\ is comparable to
the local fraction of $\sim$ 70\% \citep{kob06}, then the \GC\ contains
enough binary stars to interact with the central black hole and to produce
an observable population of HVSs.

The ejection rate from the central black hole depends on the structure
of the inner Galaxy near \Sgr.  For a random phase-space distribution 
of binaries in the \GC, \citet{hil88} first argued that the ejection 
rate could be as high as $\sim 10^{-2}$ yr$^{-1}$.  \citet{yutre03} 
refined the calculation to account for the destruction of binaries whose 
orbits take them near \Sgr.  This region of phase-space --- the loss 
cone --- is replenished by random scattering between binaries and other stars. 
If a steady state is reached, the ejection rate is $\sim 10^{-5}$ yr$^{-1}$ 
for objects with binary separations within 0.3~AU.  Recently, 
\citet{perhopale06} conclude that phase-space mixing by massive star 
clusters, molecular clouds, and perhaps intermediate mass black holes
produces a significantly larger ejection rate.  They estimate a Galactic
population of 10--200 HVSs with stellar masses of 4 \msun, consistent with 
observations \citep{broetal05, broetal06a}. 

Once these ejected stars are in the halo, the velocity distributions
derived from ongoing radial velocity surveys provide a test of the
ejection model. To make a first comparison, we use data from the halo
survey of \citet{broetal06a, broetal06b} and data for 2 HVSs from
\citet{ede05} and \citet{hir05}. The Brown et al. survey selects
candidate B-type stars from the Sloan Digital Sky Survey
\citep[SDSS;][]{ade06} and derives distances, radial velocities, and
spectral types from optical spectra acquired with the 6.5-m MMT. This
survey is complete and can detect all B-type stars at any velocity out
to $\sim 100$--150~kpc.  In addition to HVSs with radial velocities
exceeding $\sim 400$~\kms, \citet{broetal06b} find an overabundance of
B-type halo stars with radial velocities of $\sim$ 250--400 km
s$^{-1}$. Because the observations show no evidence for an infalling
population with comparable radial velocities, Brown et al. conclude
that these outliers in the velocity distribution might be low velocity
HVSs or the high velocity tail of runaway B-type stars \citep{por00}.

To compare our predictions with the observations,
Figure~\ref{fig:vvsr} shows the median radial velocity as a function
of radial position in an $\alpha=2.0$ model for 3~\Msolar\ and
4~\Msolar\ primaries with secondary star masses equal to half the
primary mass.  The radial variation in the median speed reflects the
interdependence of the ejection velocity distribution, stellar lifetime, 
and survey volume. The trend toward higher median speed at larger 
results from the finite stellar lifetime of ejected stars; slow-moving 
stars cannot reach large distances within their main-sequence lifetimes 
\citep[see also][]{BauGuaPor06}. With a sizable population of ejected 
star candidates, the velocity-radius relationship in this Figure serves 
as a fundamental test of the stellar ejection hypothesis.

The comparison of the model prediction with observations of known high
velocity halo stars (open symbols in Figure~\ref{fig:vvsr}) is
encouraging.  In general, the model predicts roughly comparable
numbers of HVSs and stars with radial velocities of 250--400 km
s$^{-1}$, as observed in current surveys.  Aside from the single
subdwarf O star outlier \citep{hir05} at $(r,~v_{med}) \sim$
$(20~\rm{kpc}, 700~\rm{kms})$, the observed velocity distribution of
HVSs also closely follows the predicted relation for 4 \msun\ primary
stars. The lone 8 \msun\ HVSs \citep{ede05} at $(r,~v_{med}) \sim$
$(55~\rm{kpc}, 550~{\rm kms})$ lies within the group of lower mass
B-type stars\footnote{We include the sdO star and the 8 \msun\ B star
for completeness.  Our predicted distribution does not include massive
B main sequence stars or sdO stars.}.  At lower velocities, the
observed distribution lies between the predicted relations for 3
\msun\ and 4 \msun\ stars. With measured spectral types consistent
with stellar masses of 3--4 \msun\ for most HVSs and most of the low
velocity sample, we conclude that the observations are in excellent
agreement with model predictions.

Although \GC\ ejections are necessary to explain the observed radial
velocities above 400~km~s$^{-1}$, objects with radial velocities of
250--400 km s$^{-1}$ could be runaway B stars from supernova-disrupted
binaries or other stellar dynamical interactions
\citep[e.g.,][]{bla61,pov67}. However, for radial velocities exceeding
$\sim$ 200~km~s$^{-1}$, the velocity distribution of runaway stars
should probably fall off more steeply than is observed
\citep{leo91,por00}. Current model predictions and statistics of halo
stars with radial velocities of 250--400 km s$^{-1}$ do not provide a
conclusive test of the runaway B star scenario. A simulation of the
observable population of runaway B stars --- similar in spirit to our
simulation of the observable population of HVSs --- would provide a
good test of this mechanism.

Despite the small number of plausibly ejected stars in the sample of
\citet{broetal06b}, these objects provide an interesting test of models
of the Galactic mass distribution.  We consider the model with the
steepest large-radius fall-off in density, corresponding to $\alpha =
2.5$ in Table~1, and the HVSs listed by \citet[][Table 1
therein]{broetal06b}, and shown here in Figure~\ref{fig:vvsr}.
None of the nine late-type B stars with distances in excess of 40~kpc 
from the \GC\ and speeds above 250~\kms\ have a speed in excess of the 
model's predicted median velocity of 752~\kms. Thus, we can 
rule out the model at a confidence level of 99.9\%.  If the objects with 
speeds less than 400~\kms\ are not ejected stars, then the confidence 
level drops to about 97\%. Including the more massive early-type HVS 
\citep{ede05}, which should have a higher ejection speed than the 
late B-type stars, raises the confidence level above 98\%.  A similar 
argument using all ten B~stars suggests that the $\alpha = 2.2$ model 
is excluded with better than 98\% confidence.  The models with shallower 
large-radius fall-off can not be ruled out in this way.

These constraints on Galactic mass models depends heavily on the
assumption that the known late-type HVSs have masses close to
4~\Msolar.  If all objects were 3~\Msolar\ stars, the predicted median
velocity for ejected stars outside of 30~kpc is 437~km~s$^{-1}$ in the
$\alpha = 2.5$ model (Table~\ref{tab:medvej}).  This median speed is
consistent with the data. However, the model predicts that 25\% of
stars should have a speed greater than about 750~km~s$^{-1}$.  All
eleven late-type B~stars shown in Figure~\ref{fig:vvsr} have speeds in
excess of 250~km~s$^{-1}$ and distances greater than 30~kpc, yet none
has a speed exceeding 750~\kms. In this case, the $\alpha = 2.5$ model
may be ruled out at the 95\% confidence level; this level is even
higher if we consider the early-type HSV as well.  Note that these
results depend mainly on the mass of the ejected primary, not on the
mass of the secondary star (see Fig.~\ref{fig:fountain}).

In making these first probes of the Galactic mass distribution with
HVSs, we note that larger samples of halo stars, globular clusters,
and dwarf galaxies currently place better constraints on the mass
distribution in the Galactic halo \citep[e.g.,][and references
therein]{bat05,deh06}. In particular, these data rule out the $\alpha
\gtrsim$ 2.2 models with higher significance than the known HVSs and
lower velocity ejected stars. However, our results show that even
small samples of HVSs are a potentially important probe of the
Galactic mass distribution to large radii.

\section{Discussion and summary}

\begin{figure}[htb]
%\centerline{\includegraphics[width=5.5in]{histo.ps}}
\centerline{\includegraphics[width=5.5in]{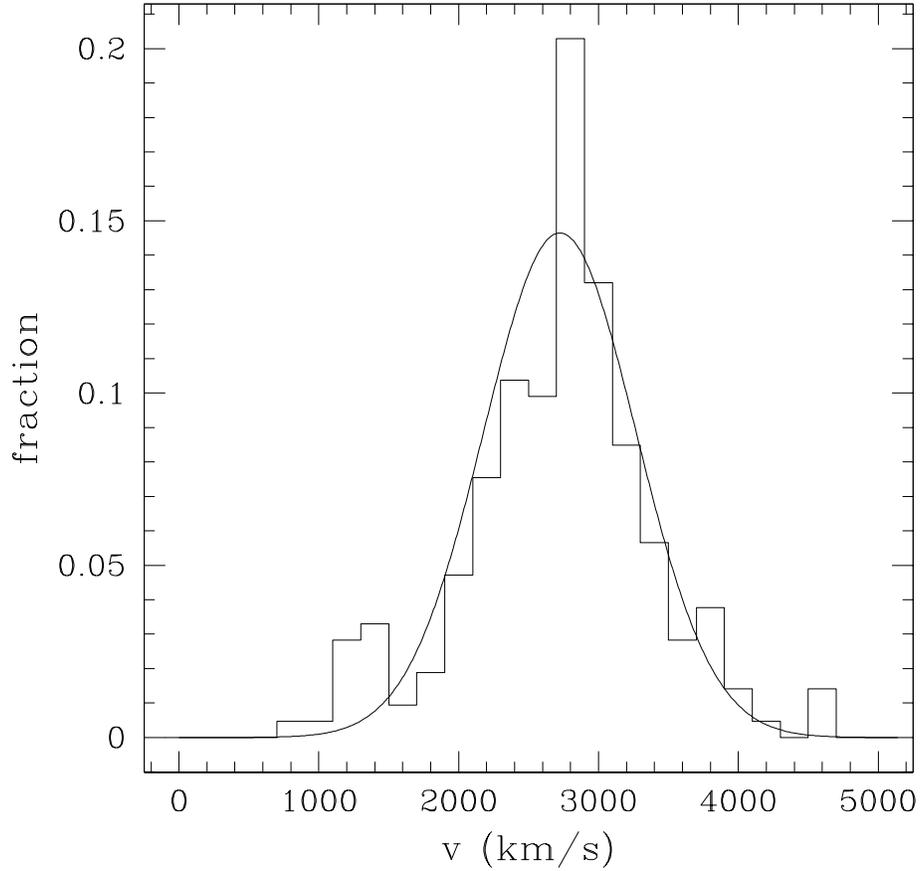}}
\caption{The distribution of ejection speeds from a binary 
with 4 \msun\ primary stars, just after encountering a black 
hole with a mass of $3.5\times 10^6$~\Msolar. The original binary
has a semimajor axis of 0.1~AU, and its center of mass is targeted
to reach a minimum distance of 5~AU from the black hole.
\label{fig:vdisp}}
\end{figure}

\begin{figure}[htb]
%\centerline{\includegraphics[width=5.5in]{fountainquad120.ps}}
\centerline{\includegraphics[width=5.5in]{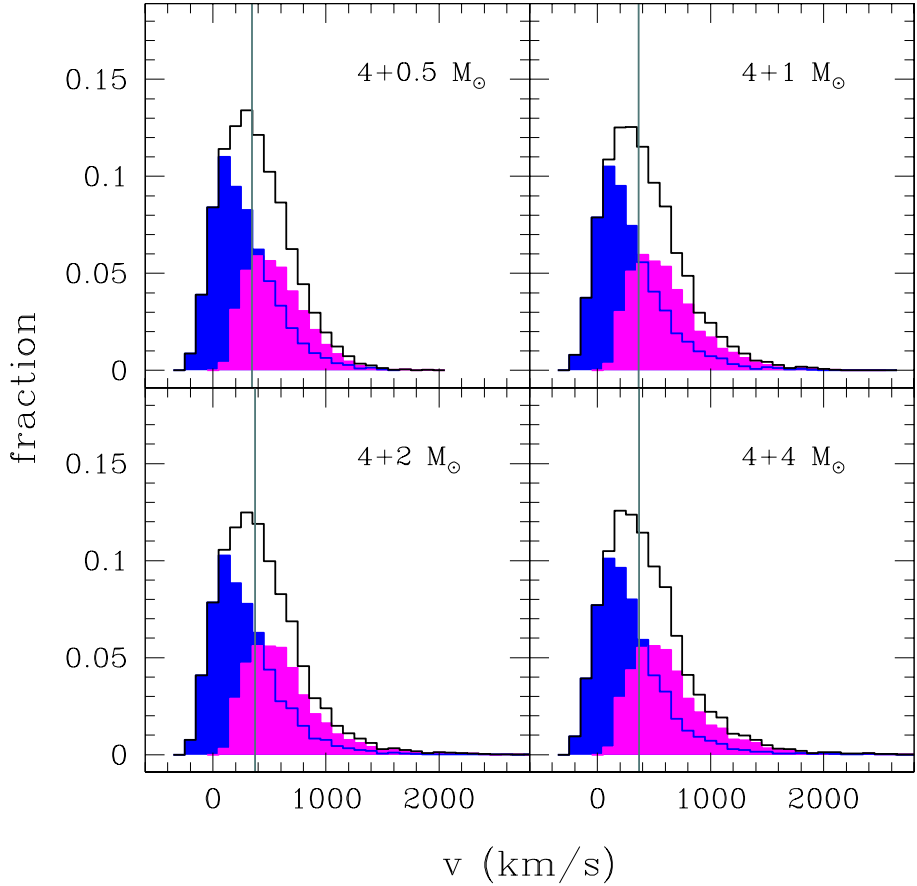}}
\caption{ The distribution of observable speeds of ejected 4 \msun\
primaries in the Galactic halo, as a function of the mass of the 
secondary. The secondary mass is indicated in
each panel.  The solid black histograms show the velocity
distribution of all objects in a survey volume which extends from
10~kpc to 80~kpc away from the \GC.  The gray vertical
lines locate the median velocity of each distribution.  The blue
histogram represents all objects inside of a 40~kpc radius; the
magenta histogram corresponds to objects outside this radius.
\label{fig:fountain}
}
\end{figure}

\begin{figure}[htb]
%\centerline{\includegraphics[width=5.5in]{fountainquada120.ps}}
\centerline{\includegraphics[width=5.5in]{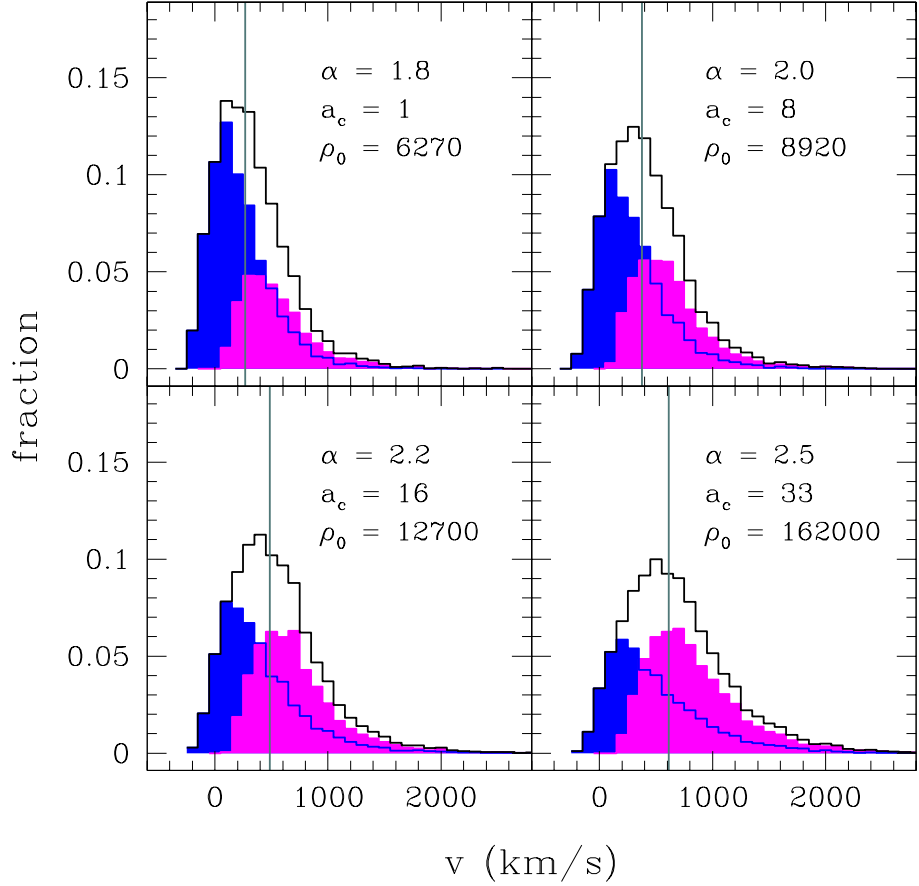}}
\caption{ Same as the previous figure, but for the distribution of
observable speeds of an ejected 4 \msun\ primary with a 2 \msun\ 
secondary, as a function of Galactic mass model. The model parameters,
displayed in each panel, correspond to eq.~(\ref{eq:massmodel}).
\label{fig:fountaina}
}
\end{figure}

\begin{figure}[htb]
%\centerline{\includegraphics[width=5.5in]{fountainquada90.ps}}
\centerline{\includegraphics[width=5.5in]{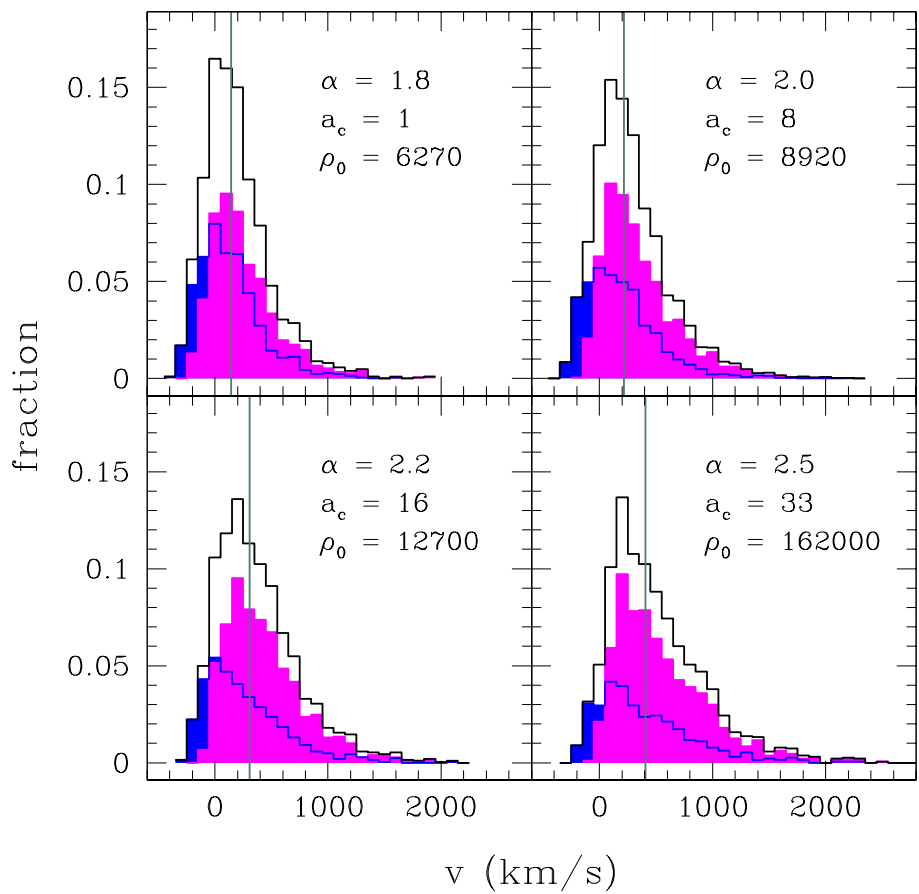}}
\caption{ Similar to the previous figure, but for the distribution of
observable speeds of an ejected  3 \msun\ primary with a 1.5 \msun\
secondary, as a function of Galactic mass model. The survey volume
in this case spans a radial extent of $10~{\rm kpc} < r < 90~{\rm kpc}$;
the blue and magenta histograms correspond to populations inside
and outside of 30~kpc, respectively.
\label{fig:fountainb}
}
\end{figure}

\begin{figure}[htb]
%\centerline{\includegraphics[width=5.5in]{vesc.ps}}
\centerline{\includegraphics[width=5.5in]{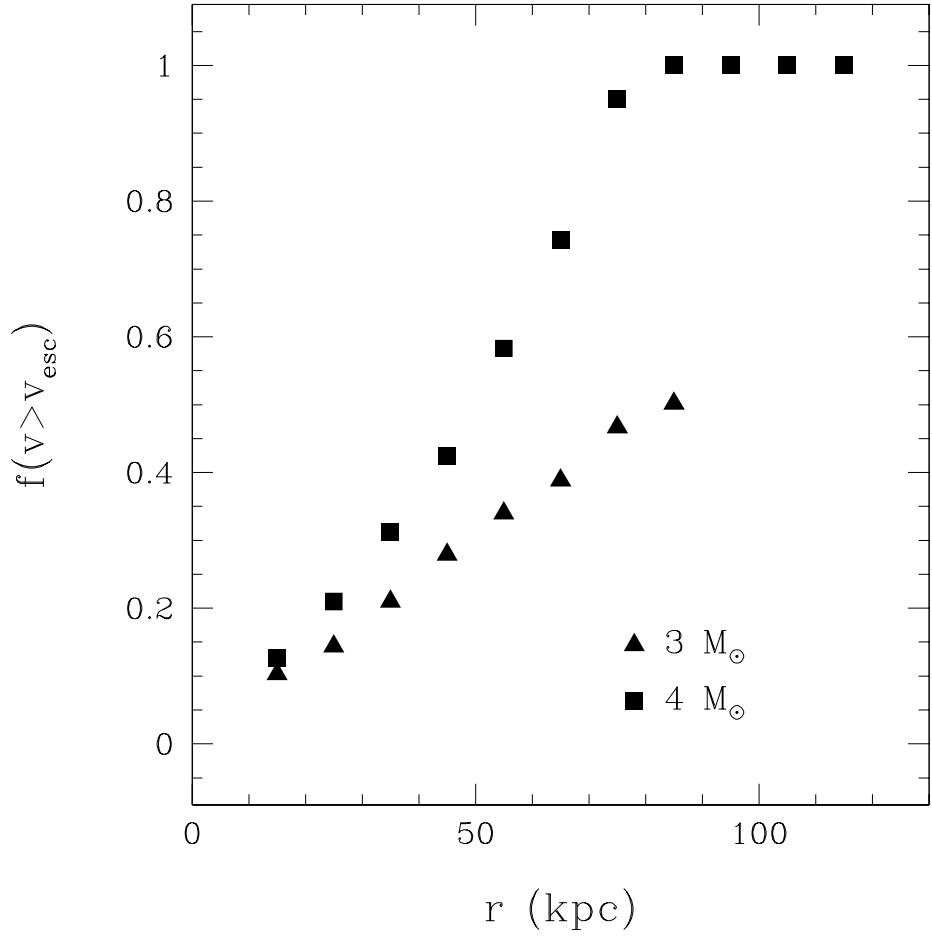}}
\caption{The fraction of observable ejected stars with radial 
velocities exceeding the escape velocity as a function of radial
distance from the \GC. As indicated in the legend, squares
show results for 4 \msun\ primary stars; triangles show
results for 3 \msun\ primaries.
\label{fig:vesc}
}
\end{figure}

\begin{figure}[htb]
%\centerline{\includegraphics[width=5.5in]{fountainvvsr.ps}}
\centerline{\includegraphics[width=5.5in]{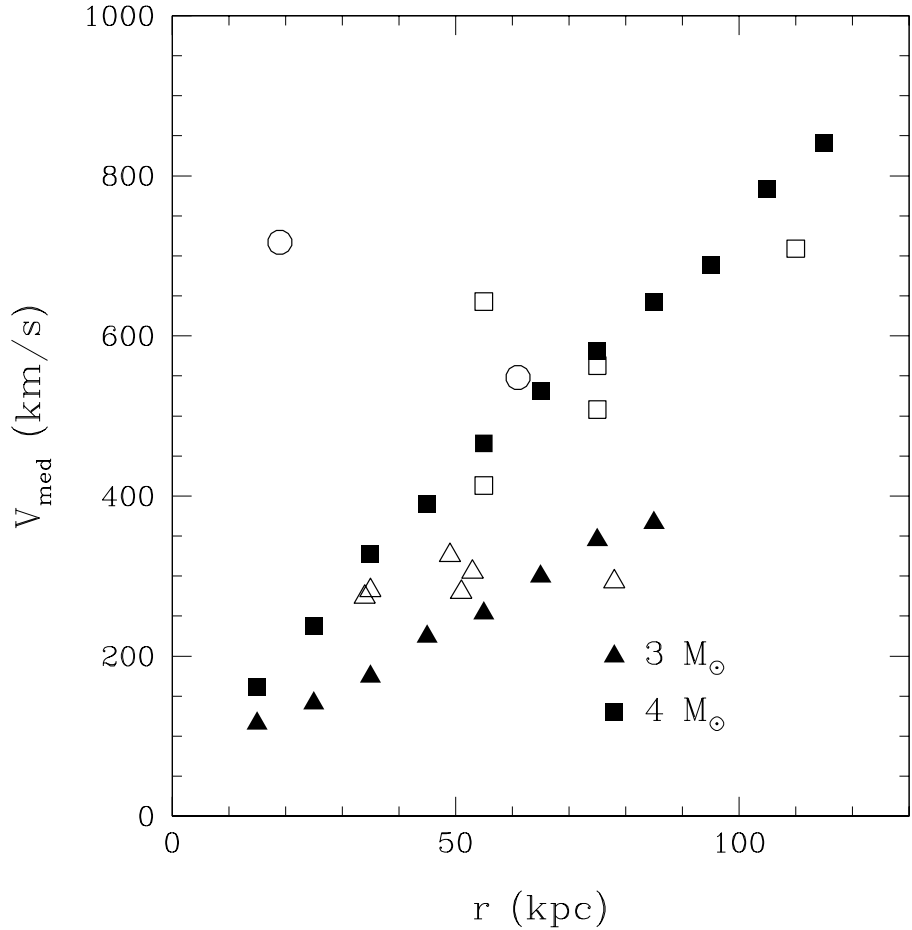}}
\caption{ The median speeds of ejected 3~\Msolar\ (filled triangles)
and 4~\Msolar\ (filled squares) primaries in the Galactic halo, as a 
function of distance from the \GC. In both cases the secondary mass is half 
that of the primary.  The open boxes are the five HVSs discovered in a 
targeted survey \citep{broetal06b}; the open triangles are lower
velocity, plausibly ejected, stars from this survey. The open circles 
are 2 HVSs discovered independently \citep{ede05,hir05}
and which are outside the mass range we consider.  \label{fig:vvsr}
}
\end{figure}

We have developed a first calculation of the predicted distribution of
radial velocities for HVSs at galactocentric distances of 10--120~kpc
in the Galactic halo.  Our simulations quantify the ejection
probabilities and velocities of binaries interacting with \Sgr,
generate the distribution of ejected stars from the vicinity of the
\GC, and produce the full phase-space radial velocity distribution of
ejected stars in the halo from an integration of the orbits of ejected
stars through the Galactic potential. In addition to providing a
framework for interpreting the observed velocity distribution of HVSs
and related stars, the results of these simulations (Figures
\ref{fig:vdisp}--\ref{fig:fountainb}) demonstrate that this approach
can provide interesting constraints on the Galactic potential and on
the mass function of the population of binary stars in the \GC.

Besides confirming Hills' (1988, 1992) results for the average
velocity of ejected stars from the black hole, we derive a broad
distribution of ejection velocities for each value of $\abin$ and
$\rclose$.  A Gaussian with $\sigma_v \sim 0.2 v_{avg}$ provides a
reasonable approximation to the derived velocity distribution.  We
plan to consider the origin of this distribution in future studies.

The initial comparisons between the model and observations are
encouraging.  The measured radial velocities of HVSs and stars with
somewhat lower speeds agree well with the predictions. The current
population of these stars provides a constraint on the Galactic mass
distribution. As ongoing radial velocity surveys yield more HVSs with
accurate stellar masses, the model will allow more robust tests for
the origin of HVSs and lower velocity stars, will provide limits on
the initial mass function of primary stars in the \GC, and will yield
better constraints on mass models for the Galaxy.

As surveys reveal more HVS, the observed velocity distribution of 
ejected stars in the Galactic halo may help to distinguish among 
alternative scenarios for producing HVSs. Compared to the single black
hole models, binary black hole models produce relatively fewer ejections 
at the highest velocities \citep{yutre03,BauGuaPor06}. Although binary 
black hole models may yield larger ejection rates overall 
\citep[e.g.,][]{yutre03}, the two models 
produce similar velocity distributions at lower velocities.  Thus, better 
constraints on the HVS distribution at the highest velocities may provide 
a test of the two pictures for HVS formation.

Refinements to the mass models can include more detailed Galactic
structure, in particular the Galactic disk.  Indeed, \citet{gneetal05}
have already proposed that HVSs can serve to probe halo
triaxiality. Including the disk potential should lead to nonradial
ejection velocities, and may cause some clustering of ejected stars in
the halo. We may therefore use the breakdown of spherical symmetry in
the distribution of ejected stars about the \GC\ to reveal the
Galaxy's structure.

If ejected stars are produced in short-lived star forming regions near
\Sgr, or if they originate in individual star clusters which settle
dynamically into the \GC, ejected stars --- including HVSs --- may
also cluster in time. Given a Galactic mass model, we can identify any
temporal correlations by using the radial position and velocity of
stars to determine the time since they were ejected.  The extraction
of temporal and spatial clustering information may be difficult with
rare HVSs. However, a clustering analysis might be feasible with the
more numerous, slower population of ejected stars suggested by our
model.

\acknowledgements

We thank an anonymous referee for raising interesting issues that 
improved the paper. 
We acknowledge support from the {\it NASA Astrophysics Theory Program} 
through grant NAG5-13278.

\begin{deluxetable}{lccccccc}
\tablecolumns{8}
\tablewidth{0pc}
\tabletypesize{\footnotesize}
\tablenum{1}
\tablecaption{Model parameters}
\tablehead{
 \colhead{parameter} & \colhead{symbol} & \colhead{value or range} 
}
\startdata
\Sgr\ mass & $\Mbh$ & 3.5$\times 10^6$ \Msolar \\
primary mass & $\ma$ & 3--4 \Msolar \\
secondary mass & $\mb$ & 0.5--4 \Msolar \\
binary semimajor axis & $\abin$ &  0.05--4 AU \\
closest approach distance & $\rclose$ &  $\sim$1--700 AU \\
Galactic density index\tablenotemark{a} & $\alpha$ & 1.8--2.5 \\
Galactic ``core'' radius & $\acore$ & 1--33 AU \\
Galactic central density\tablenotemark{b}
  & $\rho_0$ & 6270--162000 \Msolar/pc$^3$ \\
\enddata
\tablenotetext{a}{The parameters $\alpha$, $\acore$ and $\rho_0$ are
defined in eq.~(\ref{eq:massmodel})}
\tablenotetext{b}{These values do not
include the mass of \Sgr}
\label{tab:modelparms}
\end{deluxetable}

\begin{deluxetable}{lccccccc}
\tablecolumns{8}
\tablewidth{0pc}
\tabletypesize{\footnotesize}
\tablenum{2}
\tablecaption{Median observable speeds of ejected stars as a function of mass model}
\tablehead{
  \colhead{} &
  \colhead{} &
  \colhead{} &
  \multicolumn{2}{c}{3 \Msolar\tablenotemark{a}} &
  \colhead{} &
  \multicolumn{2}{c}{4 \Msolar\tablenotemark{b}} 
\\
\cline{4-5}\cline{7-8}
  \colhead{$\alpha$\tablenotemark{c}} & 
  \colhead{$\acore$ (pc)} & 
  \colhead{$\rho_0$ (\Msolar/pc$^3$)} &  
  \colhead{$v_{med}$ (full)} &
  \colhead{$v_{med}$ (halo)} &
  \colhead{} & 
  \colhead{$v_{med}$ (full)} &
  \colhead{$v_{med}$ (halo)}
}
\startdata
1.8 & 1 & $1.62\times 10^5$ & 144 & 188 & \ & 270 & 485 \\
2.0 & 8 & $1.27\times 10^4$ & 218 & 272 & \ & 374 & 565 \\
2.2 & 16 & 8920 & 306 & 370 & \ & 484 & 648 \\
2.5 & 33 & 6270 & 407 & 457 & \ & 608 & 752 \\
%\cline{1-5} \\
\enddata 
\tablenotetext{a}{ These columns list results for a primary mass of
3~\Msolar\ and a secondary mass of 1.5~\Msolar. The radial extent of
the full survey -- $v_{med}$ (full) -- is 10--90~kpc; the radial 
extent of the halo survey -- $v_{med}$ (halo) -- is 30--90~kpc.
The listed values correspond to the median radial velocity of the
ensemble of ejected stars.}
\tablenotetext{b}{ These columns list results for a primary mass of
4~\Msolar\ and a secondary mass of 2~\Msolar. The radial extent of
the full survey -- $v_{med}$ (full) -- is 10--120~kpc; the radial 
extent of the halo survey -- $v_{med}$ (halo) -- is 40--120~kpc.
The listed values correspond to the median radial velocity of the
ensemble of ejected stars.}
\tablenotetext{c}{$\alpha$, $\acore$ and $\rho_0$
are defined in eq.~(\ref{eq:massmodel}).}
\label{tab:medvej}
\end{deluxetable}

\begin{deluxetable}{cccccc}
\tablecolumns{6}
\tablewidth{0pc}
\tabletypesize{\footnotesize}
\tablenum{3}
\tablecaption{Maximum distance from the \GC\ as a function of ejected mass}
\tablehead{
  \colhead{$m_1$ ($M_{\odot}$)} & 
  \colhead{$\tlife$ (Myr)} & 
  \colhead{$m_2$ ($M_{\odot}$)} &  
  \colhead{$a_{\rm bin}$ (AU)} &  
  \colhead{$v_{ej}$ (km~s$^{-1}$)} &
  \colhead{$d_{max}$ (kpc)}
}
\startdata
7 & 43 &  2 & 0.04 & 4500 & 193 \\
7 & 43 &  4 & 0.05 & 4400 & 184 \\
7 & 43 &  7 & 0.05 & 4800 & 200 \\
9 & 26 &  2 & 0.05 & 4400  & 114 \\
9 & 26 &  4 & 0.06 & 4200  & 110 \\
9 & 26 &  9 & 0.06 & 4700  & 123 \\
12 & 16 & 3 & 0.06 & 4400 & 71 \\
12 & 16 & 6 & 0.06 & 4700 & 76 \\
12 & 16 & 12 & 0.06 & 5200 & 83 \\
15& 12 &  4 & 0.06 & 4800 &  58 \\
15& 12 &  8 & 0.07 & 4700 &  57 \\
15& 12 & 15 & 0.07 & 5200 &  62 \\
20&  8 &  5 & 0.07 & 4800 &  39 \\
20&  8 & 10 & 0.08 & 4800 &  39 \\
20&  8 & 20 & 0.09 & 5035 &  41 \\
30&  5 &  8 & 0.10 & 4700 &  23 \\
30&  5 & 15 & 0.11 & 4700 &  24 \\
30&  5 & 30 & 0.12 & 5000 &  25 \\
\enddata 
\tablenotetext{a}{These columns list the average ejection velocity
$v_{ej}$ for the specified binary semimajor axis $\abin$, and the
corresponding maximum galactocentric distance $d_{max}$ that an
ejected primary can achieve, as a function of the primary mass $m_1$
and the secondary mass $m_2$. Specifically, if the primary is ejected
with velocity $v_{ej}$, $d_{max}$ is the distance from the \GC\ the
star reaches during its main sequence lifetime $\tlife$ as it
travels radially outward through the Galactic halo. Here,
$v_{ej}$ is taken from eq.~\ref{eq:vej} with $\facR = 1$, that is,
assuming an optimal impact parameter $\rclose$. The binary
semimajor axis $\abin$ is chosen to reflect a minimum separation
as a result of Roche-lobe overflow. Note that it would be rare for 
a primary to actually achieve the listed $d_{max}$, since ejections
are not common for very tight binaries.}
\label{tab:dmax}
\end{deluxetable}

\end{document}